\documentclass{svmult}
\usepackage{graphicx}
\usepackage{makeidx}
\usepackage{multicol}

\begin{document}

\title*{Emergent Statistical Wealth Distributions in Simple Monetary Exchange Models: A Critical Review}
\titlerunning{Emergent Statistical Wealth Distributions}
\author{Thomas Lux}
\institute{Department of Economics, University of Kiel, Olshausenstr. 40, 24118 Kiel, Germany}
\maketitle

Abstract: This paper reviews recent attempts at modelling
inequality of wealth as an emergent phenomenon of
interacting-agent processes. We point out that recent models of
wealth condensation which draw their inspiration from molecular
dynamics have, in fact, reinvented a process introduced quite some
time ago by Angle (1986) in the sociological literature. We
emphasize some problematic aspects of simple wealth exchange
models and contrast them with a monetary model based on economic
principles of market mediated exchange. The paper also reports new
results on the influence of market power on the wealth
distribution in statistical equilibrium. As it turns out,
inequality increases but market power alone is not sufficient for
changing the exponential tails of simple exchange models into
Pareto tails.

\section{Introduction}
Since the days of Vilfredo Pareto, the frequency distribution of
wealth among the members of a society has been the subject of
intense empirical research. Recent research confirms that
power-law behaviour with an exponent between 1 and 2 indeed seems
to characterize the right tail of the distribution (Levy and
Solomon, 1997; Castaldi and Milakovic, 2005). However, when
applied to the entire shape of the empirical distribution, the
power law would produce a rather mediocre fit and would be
outperformed by other candidate processes like the lognormal or
Gamma distributions. As it seems to emerge from the literature, a
transition occurs in the data from an exponential
shape to power-law behavior somewhere above the 90 percent quantile again.\\\\
These and other findings should give rise to modelling efforts
explaining the remarkably similar wealth distribution of many
developed countries. Unfortunately, economic theory has been quite
silent on this topic for a long time. Until recently, one had to
go back the to literature of the fifties and sixties (e.g.,
Champernowne, 1953; Mandelbrot, 1961) to find stochastic models of
wealth accumulation in modern societies. Recent advances in
computer technology, however, open another avenue for analysis of
the emergence of wealth distributions allowing this issue to be
studied in a computational agent-based framework. Such a bottom-up
approach could, in principle, be helpful in isolating the key
mechanisms that apparently lead to a stratification of wealth in
advanced economies. As it appears, this path has been pursued
recently by physicists rather than economists (cf. Bouchaud and
M\'{e}zard, 2000; Dr\v{a}gulescu and Yakovenko, 2000; Chakraborty
and Chakrabarty, 2000; Silver, Slad and Takamoto, 2002, among
others). However, it has
 been entirely overlooked in the pertinent publications that these
models have an important predecessor in the sociological
literature. Investigating essentially the same structures already
almost twenty years ago, Angle, 1986, might be considered as the
first contribution to agent-based analysis of wealth formation. In
the following, I will shortly review Angle's interesting work as
the prototypical agent-based model of wealth dynamics, based on
particle-like microscopic interactions of agents. I will point out
aspects of this class of models (covering most of the econophysics
contributions mentioned above) that would be considered to be
problematic by economists (section \ref{sec::2}). As an
alternative framework, I will, then, review the contribution by
Silver et. al. (2002) which much better fits into standard
economic reasoning, but nevertheless provides a similarly simple
formalization of an agent-based exchange model (section
\ref{sec::3}). Section \ref{sec::4} presents some additional
results expanding on the seminal framework of Silver et. al.
Conclusions are in section \ref{sec::5}.

\section{Angle's Surplus Theory of Social Stratification and the Inequality Process}
\label{sec::2} In a long chain of papers covering more than 15
years, sociologist John Angle has elaborated on a class of
stochastic processes which he first proposed in 1986 as a
generating mechanism for the universal emergence of inequality in
wealth distributions in human societies. His starting point is
evidence he attributes to archeological excavations that
inequality among the members of a community is typically first
found with the introduction of agriculture and the ensuing prevalence of
food abundance: While simpler hunter/gatherer societies appear to
be rather egalitarian, production of a ``surplus'' beyond
subsistence level immediately seems to lead to a ``ranked
society'' or some kind of ``chiefdom'' (Angle, 1986, p. 298).

So as soon as there is some excess capacity of food, processes
seem to be set into motion from which inequality emerges. Angle,
surveying earlier narrative work in sociology, sees this as the
result of redistribution by which some members of society succeed
in grabbing some of the surplus wealth of others. The relevant empirical observations are summarized as follows:\\
\begin{center}
\parbox[1cm]{10.6cm}{
``\textbf{Proposition 1:} \emph{Where people are able to produce a surplus, some of the surplus would be fugitive
and would leave the possession of the people who produce it.}\\...\\
\textbf{Proposition 2:} \emph{Wealth confers on those who possess
it the ability to extract wealth from others. So netting out each
person's ability to do this in a general competition for surplus
wealth, the rich tend to take surplus away from the poor." (Angle,
1986, p. 298).}}
\end{center}\vspace{0.2cm}

According to Angle, the expropriation of the losers happens via
(1) theft, (2) extortion, (3) taxation, (4) exchange coerced by
unequal power between the participants, (5) genuinely voluntary
exchange, or (6) gift (\emph{ibid.}).

The process he designs as a formalisation of these ideas is a true
interacting particle model: in a finite population, agents are
randomly matched in pairs and try to catch part of the other's
wealth. A random toss $D_{t}\in\{0,1\}$ decides which of both
agents is the winner of this conflict. Angle in various papers
considers cases with equal winning probabilities $0.5$ as well as
others with probabilities being biased in favor of either the
wealthier or poorer of both individuals. If the winner of this
encounter is assumed to take away a fixed proportion of the
other's wealth, $\omega$, the simplest version of the ``inequality
process'' leads to a stochastic evolution of wealth of individuals
$i$ and $j$ who had bumped into each other according to:
\begin{eqnarray}\label{EQ1}
w_{i,t}&=&w_{i,t-1}+D_{t}\omega w_{j,t-1}-(1-D_{t})\omega w_{i,t-1},\nonumber\\\\
w_{j,t}&=&w_{j,t-1}+(1-D_{t})\omega w_{i,t-1}-D_{t}\omega
w_{j,t-1}.\nonumber
\end{eqnarray}

Time $t$ is measured in encounters and one pair of agents from the
whole population is chosen for this interaction in each period.
Angle (1986) shows via simulations that this dynamics leads to a
stationary distribution which can be reasonably well fitted by a
Gamma distribution. Angle (1993) provides an argument for why the
Gamma distribution approximates the equilibrium distribution of
the process for empirically relevant values of its parameters.
Later papers provide various extensions of the basic model. While
the exponential decay of the Gamma distribution might not be in
accordance with power law behavior at the high
 end of the richest individuals, Angle's model is the first agent-based approach matching several
 essential features of empirical wealth distributions which he carefully lists as desiderada
 (i.e. stylized facts) for a theory of inequality. Among other properties, he emphasizes the
 uni-modality with a mode above minimum income which could not be reproduced by a monotonic
 distribution function. Angle is also careful to point out that with binned data,
 realizations of his process would be hard to distinguish from realizations of Pareto random
 variables which he demonstrates via a few Monte Carlo runs.

 Unfortunately, Angle's process might be hard to accept for economists as a theory of the emergence
 of inequality in market economies.

 First, a glance at the list of the six mechanisms for appropriation of another
 agent's wealth might raise doubts about their relative importance in modern societies: for most
 countries of the world, ``theft'' should perhaps not be the most eminent mechanism for stratification
 of the wealth distribution. Note also that ``genuinely voluntary exchange'' is listed
 only at rank 5 and behind ``exchange coerced by unequal power''. However, voluntary exchange is at
 the heart of economic activity at all levels of development rather than being a minor facet.

 However, despite being mentioned in the list of mechanisms of redistribution,
 voluntary exchange is not really considered in Angle's model in which an agent simply
 takes away part of the belongings of another. What is more, this kind of encounter would - in its
 literal sense - hardly be imaginable as both agents would rather prefer \emph{not} to participate in this
 game of a burglar economy - at least if they possess a minimum degree of risk aversion. The model,
 thus, is not in harmony with the principle of voluntary participation of agents in the hypothesized process which
 economists would consider to be an important requirement for a valid theory of exchange activities. One should
 also note that another problem is the lack of consideration of the measurement of wealth
 (in terms of monetary units) and the influence of changes of the value of certain components of overall
 wealth.

 Despite these problematic features from the viewpoint of economics, Angle's model deserves credit
 as the first contribution in which inequality results as an emergent property of an agent-based
 approach. A glance at the recent econophysics literature shows that the basic building blocks of practically
  all relevant contributions share
 the structure of the inequality process formalized by equation (1). The inequality process is, for example, practically identical to the process proposed by Bouchaud and
 M\'{e}zard (2000) and isomorphic to almost all other models mentioned above. This recent
 strand of research on wealth dynamics is, therefore, almost exemplary for the lack of coordination
 among research pursued on the same topic in different disciplines and for the unfortunate duplication
 of effort that comes along with it.

 Interestingly, the above criticism concerning the structure of the exchange process had also been voiced in a
 review of monetary exchange models developed by physicists by Hayes (2002) who introduced the label of ``theft and fraud''
 economies, but restricted it to variants in which the richer could lose more (in absolute value)
 than the poor. However, it is not clear why models which introduce a certain asymmetry to avoid this
 kind of exploitation should not also suffer from the lack of willingness of agents to participate in
 their exchange processes. It, therefore, appears that one might wish to reformulate the ``burglar economies''
 in a way that brings elements of voluntary economic exchange processes into play. While the economics literature
 has not elaborated on wealth distributions emerging from exchange activities within a group of agents,
 a huge variety of approaches is available in economics that could be utilized for this purpose. An interesting
 start has been made in a recent paper by Silver, Slud and Takamoto (2002) which contains a two-good
 general equilibrium model of an economy with heterogenous agents. Somewhat ironically, the overall outcome
 of this model is the same as with the inequality process: the stationary wealth distribution turns out to be
 a Gamma distribution.

\section{An Exchange Economy with Changing Preferences}
\label{sec::3}

Unlike the framework reviewed in the previous section, the setting
of Silver et al. is an extremely familiar one for economists.
Their economy consists of two goods, denoted $x$ and $y$ which
necessitate the introduction of a relative price $p$ being defined
as the current value of a unit of good $y$ in units of good $x$.
Note that with this assumption, considerations of revaluation of
wealth components come into play which are altogether neglected in
the sociological/physical models. All agents of the economy have
their preferences formalized by a so-called Cobb-Douglas utility
function:

\begin{equation}\label{EQ2}
\begin{array}{l}
U_{i,t}=x_{i,t}^{f_{i,t}} \cdot y_{i,t}^{1-f_{i,t}}.
\end{array}
\end{equation}\

Here, $i$ and $t$ are indices for the individuals and time,
respectively. $x_{i,t}$ and $y_{i,t}$ are, therefore, the
possessions of good $x$ and $y$ by individual $i$ at time $t$ and
$f_{i,t} \in {[0,1]}$ is a preference parameter which might differ
among individuals and, for one and the same individual, might also
change over time. $U_{i,t}$, then, is utility gained by individual
$i$ at time $t$. Individuals start with a given endowment in $t=0$
and try to maximize their utility via transactions in a
competitive market where one good is exchanged against the other.
Given their possessions of both goods at some time $t-1$, it is a
simple exercise to compute their demands for goods $x$ and $y$ at
time $t$ given the current preference parameter $f_{i,t}$:
\begin{eqnarray}\label{EQ3}
x_{i,t}&=&f_{i,t}(x_{i,t-1}+p_{t}y_{i,t-1}),\nonumber\\\\
y_{i,t}&=&(1-f_{i,t})\left(\frac{x_{i,t-1}}{p_{t}}+y_{i,t-1}\right).\nonumber
\end{eqnarray}

In (\ref{EQ3}), we have used the standard assumption that agents take the price as given in a
competitive market. Note that this market, therefore, dispenses with any assumption of unequal exchange
or even exploitation which is so central to the microscopic process of the previous chapter.\\\\

Summing up demand and supply by all our agents, we can easily calculate the equilibrium price
which simultaneously clears both markets:

\begin{equation}\label{EQ4}
p_t=\frac{\sum\limits_i\left(1-f_{i,y}\right)x_{i,t-1}}{\sum\limits_i
f_{i,t}y_{i,t-1}}.
\end{equation}

After meeting in the market, each agent possesses a different bundle of goods and his wealth
can be evaluated as:

\begin{equation}\label{EQ5}
w_{i,t}=x_{i,t}+p_ty_{i,t}.
\end{equation}\

The driving force of the dynamics of the model by Silver et al. is simply the assumption of
stochastically changing preferences: all $f_{i,t}$ are drawn anew in each period independently for
all individuals. In the baseline scenario, the $f_{i,t}$ are simply drawn from a uniform distribution over
$[0,1]$, but other distributions lead to essentially the same results. The dynamics is, thus,
generated via the agents' needs to rebalance their possessions in order to satisfy their new preference ordering.
With all agents attempting to change the composition of their ``wealth'', price changes are
triggered because of fluctuations in the overall demand for $x$ and $y$. This leads to a revaluation
of agents previous possessions, $x_{i,t-1}$ and $y_{i,t-1}$, and works like a capital gain or loss.

To summarize, we have a model in which all agents are identical
except for their random preference shocks and no market or
whatsoever power is attributed to anyone. The resulting inequality
(illustrated as the benchmark case $p_{m}=0$ in Fig. 1) is,
therefore, the mere consequence of the eventualities of the
history of preference changes and ensuing exchanges of goods. We,
therefore, do not have to impose any type of ``power'' in order to
endogeneously generate a stratification of the wealth distribution
that - like the model of section 2 - is able to capture all except
the very end (the Pareto tail) of the empirical data.

\section{Some Extensions of the Monetary Exchange Model}
\label{sec::4}

The model by Silver et al. demonstrates that stratification of
wealth can result from an innocuous exchange dynamics without
agents robbing or fleecing each other. It should, therefore, be a
promising avenue to supplement the simpler dynamic models in the
previous section. In some extensions, we, therefore, tried to
explore the sensitivity of this approach to certain changes of its
underlying assumptions. Among the many sensitivity tests we could
imagine, we started with the following variations of the basic
framework:
\begin{itemize}
  \item replacement of market interaction by pairwise exchange,
  \item introduction of agents with higher bargaining power so that the outcome of pairwise matches could
        differ from a competitive framework,
  \item introduction of natural differences among agents of some kind: here we assumed that for part
        of the population, preference changes are less pronounced than for others,
  \item introduction of savings via a framework which allows for money as an additional component in the utility function.
\end{itemize}
Due to space limitations, we will not provide detailed results on all of these experiments, but will
rather confine ourselves to one particularly interesting variant: the introduction of market power.\\
Introducing market power of some sort is certainly interesting in
light of the focus of the sociological and physics-inspired
literature on issues of power of some individuals over others.
Different avenues for implementing market power seem possible.
Here, for the sake of a first exploration of this issue, we chose
a very simple and extreme one. We assume that part of the
population can act as \emph{monopolists} in pairwise encounters:
if they are matched with an agent from the complementary subset of
non-monopolists, they can demand the monopoly price. If two
non-monopolists are matched, we compute the competitive solution.
We do the same when two monopolists meet each other assuming that
their potential monopolistic power cancels out.

Although this is an almost trivial insight in economics, it should
be noted that the monopolist is not entirely free in dictating any
price/transaction combination, but has to observe the constraint
that the other agent has to voluntarily participate in the
transaction. Since the option to not agree on the transaction
would leave the monopolist with a zero gain as well, even in this
extreme market scenario ``exploitation'' is much more limited than
in a world of ``theft and fraud''. Note also that although one
could perhaps speak of exploitation (when comparing the monopoly
setting with the competitive price), no \emph{expropriation} is
involved whatsoever since even the non-monopolist will still
increase his utility by his transaction with the more ``powerful''
monopolist.

As it turns out, allowing for monopoly power indeed changes the
resulting wealth distribution. Fig. 1 shows the pdf for (fixed)
fractions of monopolists. Varying the proportion of monopolists
from 0 (the former competitive scenario with pair-wise
transactions) to 0.4 we see a slight change in the shape of the
distribution. As it happens all distributions still show
pronounced exponential decline and can be well fitted by Gamma
distributions. However, the estimated parameters of the Gamma
distribution show a systematic variation. In particular, the slope
parameter decreases with the fraction of monopolists, $p_{m}$. A
closer look at the simulation results also shows that  the average
wealth of monopolists exceeds that of other agents but the
difference decreases with increasing $p_{m}$. Note that the Gini
dispersion ratio (G) is a negative function of $\lambda$ for the
Gamma distribution:
$G=\frac{\Gamma(\lambda+0.5)}{\pi^{2}\Gamma(\lambda+1)}$, so
that the increasing inequality would also be indicated by this popular statistics.\\
\newpage
\begin{center}
\begin{figure}[h!]
\centering
  \includegraphics[width=8cm,angle=90]{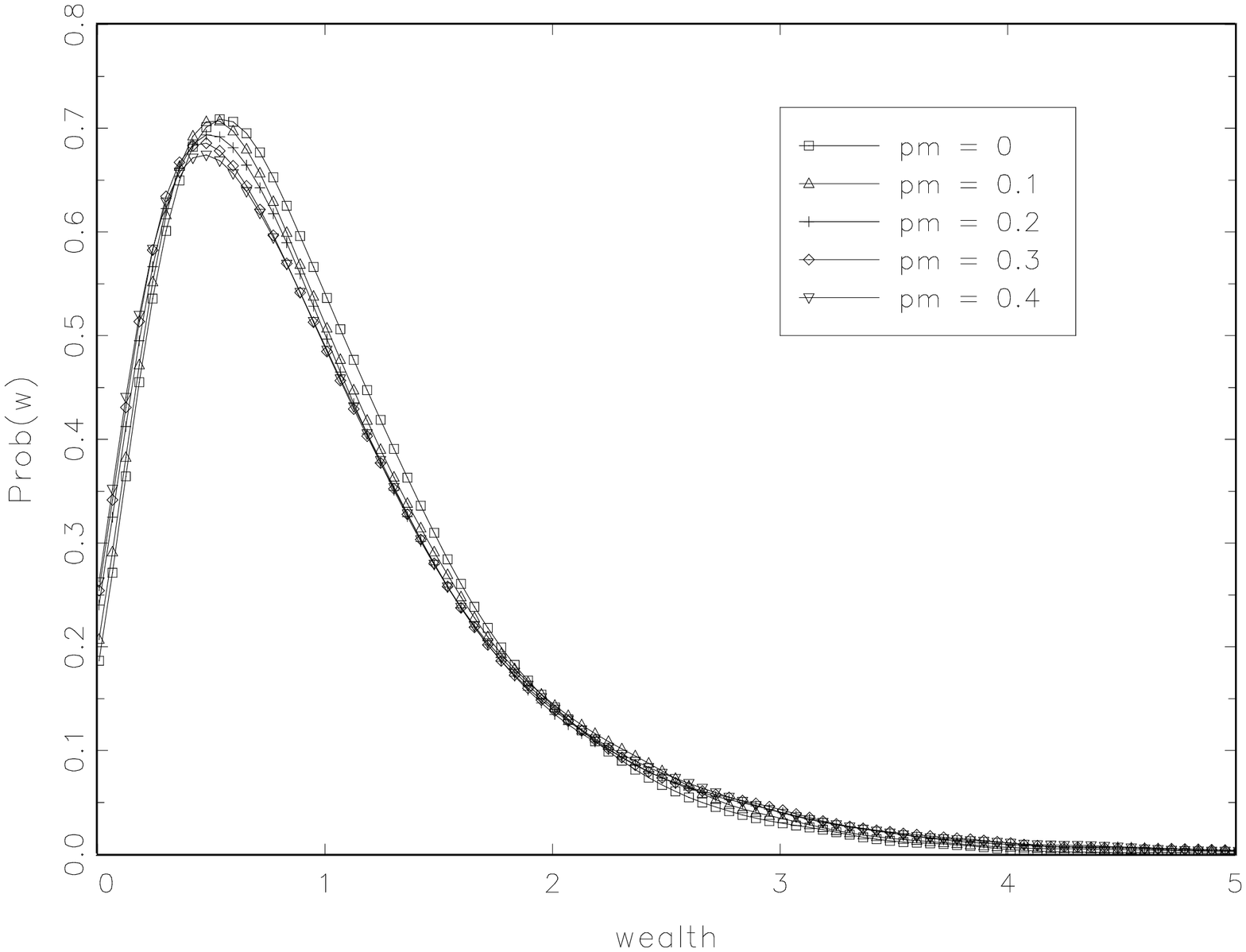}\\
\end{figure}
\end{center}
Fig. 1: Kernel estimates of statistical wealth distributions with
different fractions of monopolistic agents $p_{m}$. Results are
from simulations with 10,000 agents recorded after $5*10^{5}$
trading rounds.\\\\

The result that monopoly power is not neutral with respect to the
distribution of wealth is certainly reassuring. However, we may
also note that its introduction in the present framework does not
lead to a dramatic change of the shape of the distribution. In
particular, it does not seem to lead to anything like a Pareto
tail in place of the exponential tail of the more competitive
society. Since we have already chosen the most extreme form of
market power in the above setting it seems also unlikely that one
could obtain widely different results with milder forms of
bargaining power.

\section{Conclusions and Outlook to Future Research}\label{sec::5}

What kind of conclusions can be drawn from this review of
different approaches to agent-based models of wealth
stratification? First, it is perhaps obvious that this author
would like to advocate an approach in line with standard
principles of economic modelling. If one is not willing to follow
the emphasis of the sociological literature on all types of
exertion of power, and if one tends to the view that wealth is
influenced more by legal economic activity than by illegal theft
and fraud, economic exchange should be explicitly incorporated in
such models. This would also help to identify more clearly the
sources of the changes of wealth. Note that despite the voluntary
participation of agents in the exchange economy and the
utility-improving nature of each trade, a change in the
distribution of wealth comes with it. The difference to earlier
models is that the changes in wealth are explained by deeper,
underlying economic forces while they are simply introduced as
such in the models reviewed in sec. 2. Market exchange models also
allow to consider changes of monetary evaluation of goods and
assets as a potentially important source of changes in an
individual's nominal wealth.

Unfortunately, monetary exchange so far does not provide an
explanation of the power-law characterizing the far end of the
distribution. As we have shown above, even an extremely unequal
distribution of market power within the population seems not
sufficient to replicate this important empirical feature.
Following recent proposals in the literature one could try
additional positive feedback effects that give agents with an
already high level of wealth an additional advantage (West, 2005;
Sinha, 2005).

In the above model, one could argue that the more wealthy agents
would also acquire more bargaining power together with their
higher rank in the wealth hierarchy. Whether this would help to
explain the outer region, remains to be analyzed. However, there
are perhaps reasons to doubt that the Pareto feature might be the
mere result of clever bargaining. A glance at the Forbes list of
richest individuals (analyzed statistically by Levy and Solomon,
1997, and Castaldi and Milakovic, 2005) reveals that the upper end
of the distribution is not populated by smart dealers who in a
myriad of small deals succeeded to outwit their counterparts.
Rather, it is the founders and heirs of industrial dynasties and
successful companies operating in new branches of economic
activity whom we find there\footnote{While the majority of
entrants in the Forbes list  might fall into that category, a few
are, in fact, rather suggestive of ``theft and fraud'' avenues to
big fortunes.}. The conjecture based on this anecdotal evidence
would be that the upper end of the spectrum has its roots in risky
innovative investments. Few of these succeed but the owners behind
the succeeding ones receive an overwhelming reward. This would
suggest that models without savings and investments should lack a
mechanism for a power law tail. One would, therefore, have to go
beyond such conservative models and combine their exchange
mechanism (which works well for the greater part of the
distribution) with an economically plausible process for the
emergence of very big fortunes.\newpage
\begin{large}
\textbf{Acknowledgement}
\end{large}\\\\
I would like to thank John Angle, Mishael Milakovic and Sitabhra
Sinha for stimulating comments and discussions and Bikas
Chakrabarti for raising my interest in the issues explored in this
paper.
\\\\
\begin{large}
\textbf{References}\\\\
\end{large}
\textsc{Angle, J.}, 1986, The Surplus Theory of Social
Stratification and the Size Distribution of Personal Wealth,
\emph{Social Forces 65},
293-326.\vspace{1mm}\\
\textsc{Angle, J.}, 1992, The Inequality Process and the
Distribution of Income to Blacks and Whites, \emph{Journal of
Mathematical
Sociology 17}, 77-98.\vspace{1mm}\\
\textsc{Angle, J.}, 1993, Deriving the Size Distribution of
Personal Wealth from ``The Rich Get Richer, the Poor Get Poorer",
\emph{Journal of
Mathematical Sociology 18}, 27-46.\vspace{1mm}\\
\textsc{Angle, J.}, 1996, How the Gamma Law of Income Distribution
Appears Invariant under Aggregation, \emph{Journal of Mathematical
Sociology 31}, 325-358.\vspace{1mm}\\
\textsc{Bauchaud, J.-P. and M. M\'{e}zard}, 2000, Wealth
Condensation in a
Simple Model of Economy, \emph{Physica A 282}, 536-545.\vspace{1mm}\\
\textsc{Castaldi, C. and M. Milakovic}, 2005, Turnover Activity in
Wealth Portfolios,
Working Paper, University of Kiel.\vspace{1mm}\\
\textsc{Chakraborti, A. and B. Chakrabarti}, 2000, Statistical
Mechanics of Money: How Saving Propensities Affects its
Distribution,
\emph{European Physical Journal B 17}, 167-170.\vspace{1mm}\\
\textsc{Champernowne, D.}, 1953, A Model of Income
Distribution,\emph{
Economic Journal 53}, 318-351.\vspace{1mm}\\
\textsc{Dr\v{a}gulescu, A. and V. Yakovenko}, 2000, Statistical
Mechanics of Money, \emph{European Physical Journal B 17},
723-729.\vspace{1mm}\\
\textsc{Hayes, B.}, 2002, Follow the Money, \emph{American
Scientist 90},
2002, 400-405.\vspace{1mm}\\
\textsc{Levy, M. and S. Solomon}, 1997, New Evidence for the
Power-Law
Distribution of Wealth, \emph{Physica A 242}, 90-94.\vspace{1mm}\\
\textsc{Mandelbrot, B.}, 1961, Stable Paretian Random Functions
and the Multiplicative Variation of Income, \emph{Econometrica
29},
517-543.\vspace{1mm}\\
\textsc{Sinha, S.}, 2005, Pareto-Law Wealth Distribution in an
Asset Exchange Economy with Wealth Dependent Asymmetry, Working
Paper,
Institute of Mathematical Sciences, Chennai.\vspace{1mm}\\
\textsc{Silver, J., E. Slud and K. Takamoto}, 2002, Statistical
Equilibrium Wealth Distributions in an Exchange Economy with
Stochastic Preferences, \emph{Journal of Economic Theory 106},
417-435.\vspace{1mm}\\
\textsc{Scafetta, N., B. West and S. Picozzi}, 2003, A
Trade-Investment Model for Distribution of Wealth, cond-mat
0306579.
\end{document}